# IoT Device Identification Using Deep Learning

Jaidip Kotak and Yuval Elovici

Department of Software and Information Systems Engineering, Ben-Gurion University,
Beer-Sheva, Israel
jaidip@post.bgu.ac.il, elovici@bgu.ac.il

**Abstract.** The growing use of IoT devices in organizations has increased the number of attack vectors available to attackers due to the less secure nature of the devices. The widely adopted bring your own device (BYOD) policy which allows an employee to bring any IoT device into the workplace and attach it to an organization's network also increases the risk of attacks. In order to address this threat, organizations often implement security policies in which only the connection of white-listed IoT devices is permitted. To monitor adherence to such policies and protect their networks, organizations must be able to identify the IoT devices connected to their networks and, more specifically, to identify connected IoT devices that are not on the white-list (unknown devices). In this study, we applied deep learning on network traffic to automatically identify IoT devices connected to the network. In contrast to previous work, our approach does not require that complex feature engineering be applied on the network traffic, since we represent the "communication behavior" of IoT devices using small images built from the IoT devices' network traffic payloads. In our experiments, we trained a multiclass classifier on a publicly available dataset, successfully identifying 10 different IoT devices and the traffic of smartphones and computers, with over 99% accuracy. We also trained multiclass classifiers to detect unauthorized IoT devices connected to the network, achieving over 99% overall average detection accuracy.

## 1 Introduction

The term "Internet of Things" is defined as a group of low computing devices capable of sensing and/or actuating, abilities which extend Internet connectivity beyond that of standard devices like computers, laptops, and smartphones. The number of IoT devices connected to the Internet has already surpassed the number of humans on the planet, and by 2025, the number of devices is expected to reach around 75.44 billion worldwide [1]. IoT devices are accompanied by new vulnerabilities due to a lack of security awareness among vendors and an absence of governmental standards for IoT devices [2]. As these devices become part of existing organizational networks, they expose these networks to adversaries. By using search engines like Shodan [3], hackers can easily locate

IoT devices and target them due to their lack of security. In addition, the many IoT devices connected to the organization network may be used by adversaries during a targeted attack on the organizational network. [4, 5]. An important first step in reducing the threat such devices pose and increasing security is to identify the IoT devices connected to the organizational network.

It is a challenge for organizations to identify the various IoT devices connected to their networks. Policies like the popular bring your own device (BYOD) policy exacerbate this, as employees can bring their devices into the workplace, any of which might pose a threat when connected to the organizational network [6, 7, 8]. Moreover, as different IoT devices use different protocols to communicate with other devices and/or to their respective servers, it is difficult to both maintain the security of the devices and perform post-incident investigations using traditional methodologies [9, 10]. In order to address these challenges, organizations need a means of identifying the IoT devices (both known and unknown to network administrators) connected to their networks; the ability to do so will enable organizations to more effectively handle IoT device security issues and determine whether the behavior of the connected IoT devices is normal.

Previous research proposed ways of identifying IoT devices by analyzing the network traffic [13, 14, 15]. Because existing methods are based on machine learning techniques, they require feature engineering, i.e., extraction, selection, and tuning of the features. This requires manual input from domain experts, which is both prone to errors and expensive. Existing approaches require multiple sessions to identify known and unauthorized IoT devices and hence require more time. As they are based on a multi-stage model, the architecture is more complex. Our approach addresses these limitations; it has a simple architecture, requires a single session to detect and identify known and unauthorized IoT devices, and is free from the overhead of feature engineering and the errors that may be added during feature engineering.

In this paper, we propose an approach that allows us to identify known IoT devices in the network; by using the same approach, we can also identify the presence of any unknown IoT devices in the network, as shown in our second experiment. It is easy to spoof MAC addresses, and the use of DHCP by organizations makes it difficult to identify IoT devices in the network using traditional approaches [11, 12]. Therefore, our approach is focused on the TCP content of the packets exchanged by devices as opposed to the header of the packets. While other research [13, 14, 15] has proposed methods aimed at tackling the problem of identifying IoT devices in organizational networks, our approach provides a more generic and less complex solution with accuracy comparable or greater than that of existing approaches.

The contributions of our research are as follows:
- To the best of our knowledge, we are the first to apply deep learning techniques on the TCP payload of network traffic for IoT device classification and identification.

43- Our approach can be used for the detection of white-listed IoT devices in the network traffic.
- When using our approach, only a single TCP session is needed to detect the source IoT device, in contrast to existing approaches which require multiple TCP sessions to detect the source IoT device.
- Our approach is simple in terms of architecture and free from feature engineering overhead.

## 2  Related Work

Machine learning has been used by researchers to classify network traffic for the purpose of identifying the services being used on the source computers [16, 17]. Transfer learning techniques have been used for network traffic classification, with promising results [18]. The use of machine learning and deep learning algorithms to detect malicious and benign traffic has also been demonstrated [19, 20]. Another study [21] examined machine learning techniques that can be used by adversaries to automatically identify user activities based on profiling the traffic of smart home IoT device communications.

In [13], the authors proposed a machine learning approach to identify IoT traffic based on the features of a TCP session. Each TCP session was represented by a vector of features from the network, transport, and application layers, and other features were added based on publicly available data, such as Alexa Rank [22] and GeoIP [23]. In this study, nine different IoT devices were classified based on different machine learning models using 33,468 data instances. For each classifier, an optimal threshold value was obtained, which helped identify which class the traffic belonged to. In addition, for each IoT device class, a threshold value for the number of sequences of TCP sessions was obtained, enabling the authors to determine the IoT device class for any input session. Although high accuracy (over 99%) was achieved for the task of correctly identifying the IoT device, the study had limitations in that the features selected from the application layer were limited to HTTP and TLS protocols only, and there is a need for various types of machine learning models and different numbers of TCP session sequences (threshold values) in order to identify different IoT devices.

Sivanathan *et al.*[15] characterized the IoT traffic based on the statistical attributes, such as port numbers, activity cycles, signaling patterns, and cipher suites. They proposed a multistage machine learning model to classify the IoT devices, using flow level features like the flow volume, flow duration, flow rate, sleep time, DNS interval, and NTP interval extracted from network traffic. In this experiment, the authors used 50,378 labeled instances to classify 28 IoT devices and one non-IoT device class and obtained over 99% accuracy. A limitation of this work was its complex multistage architecture and the need for a subject matter expert to decide the features to be used. Dependency on features such as the port number and domains can be risky as they can easily be altered by vendors.

Meidan *et al.*[14] applied machine learning techniques to detect white-listed IoT devices in the network traffic. Ten IoT devices were considered in this study, and in each of the experiments performed, one IoT device was removed from the white-list and treated as an unknown IoT device. More than 300 features were used to train the model, including features from different network layers and statistical features. They found that the 10 features most heavily influencing classification were mainly statistical features derived from the time to live (TTL) value. In this study, white-listed IoT device types were classified to their specific types with an average accuracy of 99%. As this approach is based on the work of [13], it inherits its limitations (as described above).

## 3 Proposed Methodology

### 3.1 Approach

The two main approaches used in the domain of network traffic classification are the rule-based approach and an approach based on statistical and behavioral features [24, 25]. The rule-based approach focuses on the port number of the service, which cannot be relied upon today. Statistical and behavioral-based features also have limitations, including the need to identify the correct features and preprocess those features to feed into the machine learning model, both of which require significant domain knowledge.

In this paper, we propose a novel approach which is based on representation learning. We conducted two experiments, one aimed at identifying IoT devices in the network traffic and the other aimed at identifying connected IoT devices that are not on the white-list (unknown devices) in the organizational network. The intuition behind this work is the small length and particular patterns of data transferred from IoT devices compared to that of computers and smartphones which use different protocols and have variable data lengths. The scope of this research is limited to IoT devices that utilize the TCP protocol for communication. However, we believe this method can also be applied to IoT devices that communicate using other protocols, including Bluetooth, ZigBee, CoAP, and others. Our research addresses the limitations of the previous research performed by other researchers (as described in the previous section).

### 3.2 Data Preprocessing

In the preprocessing phase, we convert the network traffic available in pcap (packet capture) format to grayscale images. Our focus is on the payloads of the TCP sessions which are exchanged between IoT devices, as shown in Figure 1. The data processing step is same for both of the experiments.

The TCP session payloads are converted to images using the following steps:

**Step 1:** In this step, multiple pcap files are created from a single large pcap file based on the sessions (i.e., a group of packets with the same source and destination IP address, source and destination port number, and protocol belong to a single session). The output is comprised of multiple pcap files where each pcap file represents a single session. A tool like SplitCap [26] can be utilized to perform this step. As our scope is limited to the TCP protocol, pcap files with UDP session data are removed.

**Step 2:** After obtaining the pcap files (as described above in step 1), the files are divided into groups based on the source MAC address. At the end of this step, we will have multiple folders, each of which contains the pcap files that originated from a particular MAC address. We then identify the folders that contain the traffic of the IoT devices used in our experiments based on the source MAC addresses.

**Step 3:** We remove the header of each of the packets present in a single pcap file and convert it into a bin file (binary file) which will contain the content of the TCP payload in the form of hexadecimal values. Files with no data will also get generated for cases in which there was no communication in the TCP session; in this step, these files will be removed, along with any duplicate files.

**Step 4:** We adjust the file size so that each bin file is 784 bytes; if the file size is more than 784 bytes, we trim it, and we pad files that are less than 784 bytes with 0x00 bytes.

**Step 5:** This is an optional step where bin files are converted to 28 × 28 pixel grayscale images where each pixel represents a two hexadecimal number (in total representing 1,568 hexadecimal numbers in 784 bytes) as shown in Figure 1. The images generated for each class are converted to IDX files which are similar to MNIST dataset files [27] for ease when training the deep learning model.

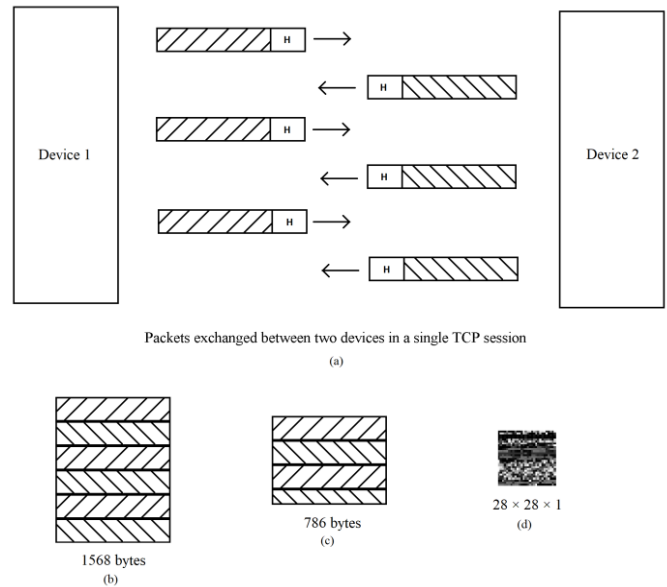

**Fig. 1.** Converting network traffic into image representation

Figure 2 presents four random images generated based on the above steps for a Belkin Wemo motion sensor and Amazon Echo. From the images, it is evident that each IoT device has a distinct pattern of communication when compared to other IoT devices.

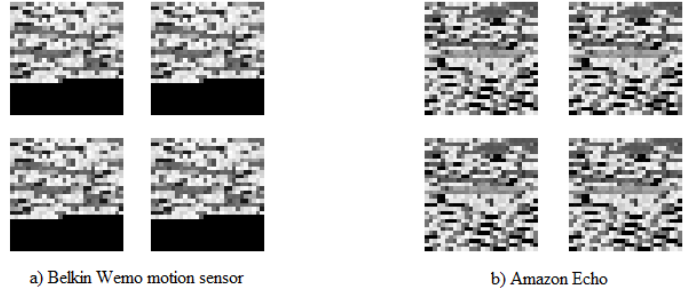

**Fig. 2.** Visualization of the communication patterns of two different IoT devices

### 3.3 Dataset and Environment

In order to verify the effectiveness of the proposed model, we trained a single layer fully connected neural network. The dataset used in our experiments came from the IoT Trace Dataset [15]. The complete dataset contains 218,657 TCP sessions, 110,849 TCP sessions from IoT devices and 107,808 TCP sessions from non-IoT devices. As our approach is based on deep learning, we only considered devices with over 1,000 TCP sessions. We divided the reduced dataset into three sets, i.e., training, validation, and test sets. In both the validation and test set, 10% of the data was randomly chosen. Table 1 contains details on the subset of the IoT Trace dataset used in our experiments (after the above mentioned preprocessing steps were performed).

**Table 1.** Dataset used in our experiments

| Device Name | MAC Address | Device Type | Sessions in Training Set | Sessions in Validation Set | Sessions in Test Set | Total Sessions |
|---|---|---|---|---|---|---|
| Belkin Wemo motion sensor | ec:1a:59:83:28:11 | IoT | 7313 | 903 | 813 | 9029 |
| Amazon Echo | 44:65:0d:56:cc:d3 | IoT | 2903 | 358 | 323 | 3584 |
| Samsung SmartCam | 00:16:6c:ab:6b:88 | IoT | 3285 | 405 | 365 | 4055 |
| Belkin Wemo switch | ec:1a:59:79:f4:89 | IoT | 2759 | 341 | 307 | 3407 |
| Netatmo Welcome | 70:ee:50:18:34:43 | IoT | 1894 | 234 | 210 | 2338 |
| Insteon camera | 00:62:6e:51:27:2e | IoT | 2177 | 269 | 242 | 2688 |
| Withings Aura smart sleep sensor | 00:24:e4:20:28:c6 | IoT | 906 | 112 | 100 | 1118 |
| Netatmo weather station | 70:ee:50:03:b8:ac | IoT | 5695 | 703 | 633 | 7031 |
| PIX-STAR photoframe | e0:76:d0:33:bb:85 | IoT | 31199 | 3852 | 3467 | 38518 |
| Non-IoT devices | N/A | Non - IoT | 20035 | 2474 | 2226 | 24735 |

### 3.4 Model Architecture

In our proposed model, we have only an input layer and an output layer. This makes our approach less complex than existing approaches for classifying IoT devices. The input to the model in both experiments is a 28 × 28 pixel (i.e., 784 pixel value) grayscale image, or if byte values are directly supplied, the input layer has 784 neurons. As we are classifying nine IoT devices along with one non-IoT device class in experiment one, the output layer will have ten neurons, as shown in Figure 3. In our second experiment, which is aimed at detecting unauthorized IoT devices (which is not part of white-list), we focus on the traffic of nine IoT devices and train nine models with similar architectures, keeping one class of IoT traffic out of the training set each time and treating the excluded IoT class traffic as an unknown IoT device to demonstrate the feasibility of our approach. Hence, in the output layer of the second experiment we have eight neurons.

For the input and output layers we initialized weights using a *normal* distribution. The activation function used in the input layer is *ReLU,* and the activation function in the output layer is *softmax*. We used the *Adam* optimizer, along with *categorical cross-entropy* as the loss and *accuracy* as the evaluation metric (for the validation set) [28, 29, 30, 31, 32]. We also validated the results by adding intermediate hidden layers with different parameters, but the results were more or less the same.

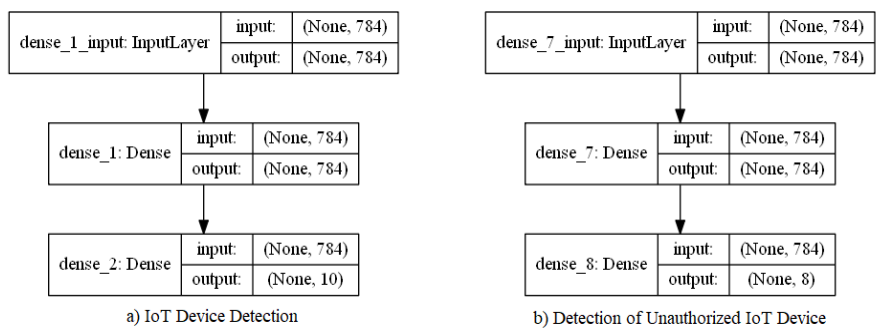

**Fig. 3.** Neural network model architecture

### 3.5 Evaluation Matrix

To evaluate our model on the test set we used: accuracy (A), precision (P), recall (R), and the F1 score (F1) as our evaluation matrix. The formula used to calculate each of these parameters is shown in Equation 1, where the TP, TN, FP, and FN are respectively the true positive, true negative, false positive, and false negative.

$$A = \frac{TP + TN}{TP + FP + FN + TN}, \qquad P = \frac{TP}{TP + FP}$$

$$R = \frac{TP}{TP + FN}, \qquad F1 = \frac{2 \times P \times R}{P + R}$$

(1)

## 4 Evaluation

**Experiment 1: Detection of IoT Devices**

Figure 4 presents the model accuracy and loss obtained on the validation set after training a multiclass classifier for 25 epochs, with a batch size of 100. Based on the graphs, it is clear that the optimum accuracy (99.87%) and optimum loss were obtained after seven epochs. Hence, we retrained the model (to avoid overfitting) for seven epochs and found that the accuracy for the test set was 99.86%, which is greater than or equal to that of existing approaches. The confusion matrix for each IoT device and the non-IoT device is presented in Table 2.

**Table 2.** Test set's confusion matrix

| Actual IoT device/classified as | 0 | 1 | 2 | 3 | 4 | 5 | 6 | 7 | 8 | 9 |
|---|---|---|---|---|---|---|---|---|---|---|
| 0- Non-IoT devices | 2226 | 0 | 0 | 0 | 0 | 0 | 0 | 0 | 0 | 0 |
| 1- Amazon Echo | 1 | 812 | 0 | 0 | 0 | 0 | 0 | 0 | 0 | 0 |
| 2- Samsung SmartCam | 2 | 0 | 321 | 0 | 0 | 0 | 0 | 0 | 0 | 0 |
| 3- Belkin Wemo switch | 0 | 0 | 0 | 365 | 0 | 0 | 0 | 0 | 0 | 0 |
| 4- Netatmo Welcome | 0 | 0 | 0 | 0 | 306 | 1 | 0 | 0 | 0 | 0 |
| 5- Insteon camera | 0 | 0 | 0 | 0 | 0 | 210 | 0 | 0 | 0 | 0 |
| 6- Withings Aura smart sleep sensor | 1 | 0 | 0 | 0 | 0 | 0 | 241 | 0 | 0 | 0 |
| 7- Netatmo weather station | 0 | 0 | 0 | 0 | 0 | 0 | 0 | 100 | 0 | 0 |
| 8- PIX-STAR photoframe | 0 | 0 | 0 | 0 | 0 | 0 | 0 | 0 | 628 | 5 |
| 9- Belkin Wemo motion sensor | 1 | 0 | 0 | 0 | 0 | 0 | 0 | 0 | 1 | 3465 |

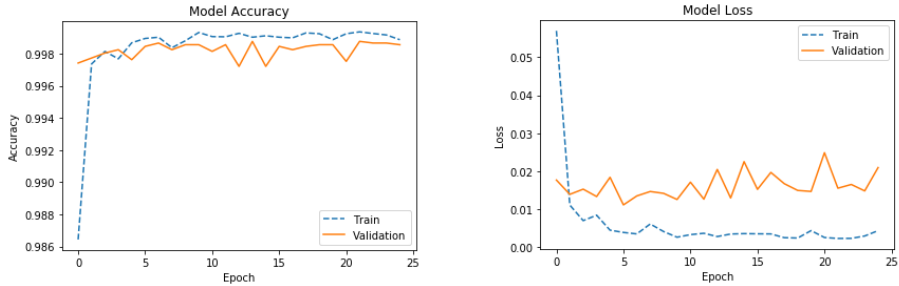

**Fig. 4.** Model accuracy and loss for validation set

**Experiment 2: Detection of Unauthorized IoT Devices**
For this experiment, the dataset included the traffic of nine IoT devices. We trained nine different multiclass classifiers with the architecture shown in Figure 3(b); each time, eight classes were used for training classifier, excluding one class which was considered an unknown IoT device. We determined the minimum number of epochs required by each classifier to obtain the maximum accuracy on the validation set.

**Table 3.** Minimum number of epochs required to obtain maximum accuracy on the validation set

| Unknown Devices | Minimum Number of Epochs | Thresholds | Test Set Accuracy (%) |
| --- | --- | --- | --- |
| Amazon Echo | 9 | 0.97 | 98.9 |
| Samsung SmartCam | 27 | 0.99 | 97.9 |
| Belkin Wemo switch | 5 | 0.77 | 99.3 |
| Netatmo Welcome | 18 | 0.99 | 98.3 |
| Insteon camera | 8 | 0.92 | 98.8 |
| Withings Aura smart sleep sensor | 6 | 0.80 | 99.8 |
| Netatmo weather station | 3 | 0.76 | 99.8 |
| PIX-STAR photoframe | 3 | 0.87 | 99.8 |
| Belkin Wemo motion sensor | 3 | 0.90 | 99.0 |

After determining the minimum number of epochs for each classifier, the classifiers were retrained using their respective minimum number of epochs. When applied to a single instance, each classifier outputs a vector of posterior probabilities with a length of eight. Each probability value denotes the likelihood of the instance to originate from one of the eight IoT devices. We used the threshold value to derive the classification of an instance, such that given the vector of probabilities, if any single probability that exceeds the threshold value exists, the instance is classified as originating from one of the eight IoT devices, depending on the index of the probability in the output vector; otherwise, that instance is classified as unknown. The threshold values were derived using the validation set on the classifiers trained, which maximizes accuracy (A) (see Equation 1). The threshold values for each classifier are described in Table 3. Higher threshold values indicate that the classifier is able to bifurcate unknown devices with greater confidence. A confusion matrix for each classifier is presented in Appendix A.

## 5      Conclusion

In this research, we presented an approach that uses deep learning to identify both known and unauthorized IoT devices in the network traffic, identifying 10 different IoT devices and the traffic of smartphones and computers, with over 99% accuracy, and achieving over 99% overall average accuracy to detect unauthorized IoT devices connected to the network. We also demonstrated some of the advantages of our approach, including its simplicity (compared to existing approaches) and the fact that it requires no feature engineering (eliminating the associated overhead). The proposed approach is also generic as it focuses on the network traffic payload of the different IoT devices as opposed to the header of the packets; thus our method is applicable to any IoT device, regardless of the protocol used for communication. In future research, we plan to explore applications of our approach to additional scenarios, possibly including different network protocols that do not use a TCP/IP network stack for communication.

1111. Xiao, L., Wan, X., Lu, X., Zhang, Y., Wu, D.: IoT security techniques based on machine learning: How do IoT devices use AI to enhance security?IEEE Signal Processing Magazine, 35(5), 41-49. (2018)
12. Ling, Z., Luo, J., Xu, Y., Gao, C., Wu, K., Fu, X.: Security vulnerabilities of internet of things: A case study of the smart plug system. IEEE Internet of Things Journal, 4(6), 1899-1909.(2017)
13. Meidan, Y., Bohadana, M., Shabtai, A., Guarnizo, J. D., Ochoa, M., Tippenhauer, N. O., Elovici, Y.: ProfilIoT: a machine learning approach for IoT device identification based on network traffic analysis. In Proceedings of the symposium on applied computing,pp. 506-509 (2017).
14. Meidan, Y., Bohadana, M., Shabtai, A., Ochoa, M., Tippenhauer, N. O., Guarnizo, J. D., Elovici, Y.: Detection of unauthorized IoT devices using machine learning techniques. (2017)arXiv preprint arXiv:1709.04647.
15. Sivanathan, A., Gharakheili, H. H., Loi, F., Radford, A., Wijenayake, C., Vishwanath, A., Sivaraman, V.: Classifying IoT devices in smart environments using network traffic characteristics: IEEE Transactions on Mobile Computing, 18(8), pp.1745-1759. (2018)
16. Wang, Z.: The applications of deep learning on traffic identification. BlackHat USA, 24(11), 1-10. (2015)
17. Lopez-Martin, M., Carro, B., Sanchez-Esguevillas, A., Lloret, J.: Network traffic classifier with convolutional and recurrent neural networks for Internet of Things. IEEE Access, 5, 18042-18050.(2017)
18. Sun, G., Liang, L., Chen, T., Xiao, F., Lang, F.: Network traffic classification based on transfer learning.: Computers & electrical engineering, 69, 920-927.(2018).
19. Wang, W., Zhu, M., Zeng, X., Ye, X., Sheng, Y.: Malware traffic classification using convolutional neural network for representation learning. In: International Conference on Information Networking (ICOIN), pp. 712-717. (2017)
20. Celik, Z. B., Walls, R. J., McDaniel, P., Swami, A.: Malware traffic detection using tamper resistant features. In MILCOM - IEEE Military Communications Conference (pp. 330-335). (2015)
21. Acar, A., Fereidooni, H., Abera, T., Sikder, A. K., Miettinen, M., Aksu, H., ... Uluagac, A. S.: Peek-a-Boo: I see your smart home activities, even encrypted!(2018)arXiv preprint arXiv:1808.02741
22. Alexa top sites,http://www.alexa.com/topsites, last accessed 2020/01/26
23. Geoip lookup service,http://geoip.com/, last accessed 2020/01/26
24. Nguyen, T. T., Armitage, G.: A survey of techniques for internet traffic classification using machine learning. IEEE communications surveys & tutorials, 10(4), 56-76. (2008)
25. Zhang, J., Chen, X., Xiang, Y., Zhou, W., Wu, J.: Robust network traffic classification. IEEE/ACM transactions on networking, 23(4), 1257-1270. (2014)
26. SplitCap,https://www.netresec.com/?page=SplitCap, last accessed 2020/01/26
27. THE MNIST DATABASE of handwritten digits, http://yann.lecun.com/exdb/mnist/ last accessed, 2020/01/26
28. Usage of initializers,https://keras.io/initializers/, last accessed 2020/01/26
29. Usage of activations,https://keras.io/activations/, last accessed 2020/01/26
30. Usage of optimizers,https://keras.io/optimizers/, last accessed 2020/01/26
31. Usage of loss functions,https://keras.io/losses/, last accessed 2020/01/26
32. Usage of metrics,https://keras.io/metrics/, last accessed 2020/01/26

**Appendix A: Confusion matrix for the detection of unauthorized IoT devices**

| Actual IoT device/classified as | 1 (U) | 2 | 3 | 4 | 5 | 6 | 7 | 8 | 9 | Precision | Recall | F1-Score |
|---|---|---|---|---|---|---|---|---|---|---|---|---|
| 1- Amazon Echo (unknown) | 802 | 4 | 0 | 0 | 0 | 0 | 0 | 0 | 7 | 0.93 | 0.986 | 0.958 |
| 2- Samsung SmartCam | 0 | 323 | 0 | 0 | 0 | 0 | 0 | 0 | 0 | 0.988 | 1 | 0.994 |
| 3- Belkin Wemo switch | 0 | 0 | 365 | 0 | 0 | 0 | 0 | 0 | 0 | 1 | 1 | 1 |
| 4- Netatmo Welcome | 1 | 0 | 0 | 306 | 0 | 0 | 0 | 0 | 0 | 1 | 0.997 | 0.998 |
| 5- Insteon camera | 0 | 0 | 0 | 0 | 210 | 0 | 0 | 0 | 0 | 1 | 1 | 1 |
| 6- Withings Aura smart sleep sensor | 0 | 0 | 0 | 0 | 0 | 242 | 0 | 0 | 0 | 1 | 1 | 1 |
| 7- Netatmo weather station | 0 | 0 | 0 | 0 | 0 | 0 | 100 | 0 | 0 | 1 | 1 | 1 |
| 8- PIX-STAR photoframe | 7 | 0 | 0 | 0 | 0 | 0 | 0 | 626 | 0 | 1 | 0.989 | 0.994 |
| 9- Belkin Wemo motion sensor | 52 | 0 | 0 | 0 | 0 | 0 | 0 | 0 | 3415 | 0.998 | 0.985 | 0.991 |
| | | | | | | | | | **Weighted Avg** | **0.99** | **0.989** | **0.989** |

| Actual IoT device/classified as | 1 | 2 (U) | 3 | 4 | 5 | 6 | 7 | 8 | 9 | Precision | Recall | F1-Score |
|---|---|---|---|---|---|---|---|---|---|---|---|---|
| 1- Amazon Echo | 812 | 1 | 0 | 0 | 0 | 0 | 0 | 0 | 0 | 1 | 0.999 | 0.999 |
| 2- Samsung SmartCam (unknown) | 0 | 250 | 0 | 0 | 0 | 0 | 0 | 0 | 73 | 0.804 | 0.774 | 0.789 |
| 3- Belkin Wemo switch | 0 | 0 | 365 | 0 | 0 | 0 | 0 | 0 | 0 | 1 | 1 | 1 |
| 4- Netatmo Welcome | 0 | 1 | 0 | 306 | 0 | 0 | 0 | 0 | 0 | 1 | 0.997 | 0.998 |
| 5- Insteon camera | 0 | 0 | 0 | 0 | 210 | 0 | 0 | 0 | 0 | 1 | 1 | 1 |
| 6- Withings Aura smart sleep sensor | 0 | 0 | 0 | 0 | 0 | 242 | 0 | 0 | 0 | 1 | 1 | 1 |
| 7- Netatmo weather station | 0 | 0 | 0 | 0 | 0 | 0 | 100 | 0 | 0 | 1 | 1 | 1 |
| 8- PIX-STAR photoframe | 0 | 7 | 0 | 0 | 0 | 0 | 0 | 626 | 0 | 1 | 0.989 | 0.994 |
| 9- Belkin Wemo motion sensor | 0 | 52 | 0 | 0 | 0 | 0 | 0 | 0 | 3415 | 0.979 | 0.985 | 0.982 |
| | | | | | | | | | **Weighted Avg** | **0.979** | **0.979** | **0.979** |

| Actual IoT device/classified as | 1 | 2 | 3 (U) | 4 | 5 | 6 | 7 | 8 | 9 | Precision | Recall | F1-Score |
|---|---|---|---|---|---|---|---|---|---|---|---|---|
| 1- Amazon Echo | 811 | 0 | 1 | 0 | 0 | 0 | 0 | 0 | 1 | 1 | 0.998 | 0.999 |
| 2- Samsung SmartCam | 0 | 322 | 1 | 0 | 0 | 0 | 0 | 0 | 0 | 0.994 | 0.997 | 0.995 |
| 3- Belkin Wemo switch (unknown) | 0 | 2 | 363 | 0 | 0 | 0 | 0 | 0 | 0 | 0.894 | 0.995 | 0.942 |
| 4- Netatmo Welcome | 0 | 0 | 1 | 306 | 0 | 0 | 0 | 0 | 0 | 1 | 0.997 | 0.998 |
| 5- Insteon camera | 0 | 0 | 0 | 0 | 210 | 0 | 0 | 0 | 0 | 1 | 1 | 1 |
| 6- Withings Aura smart sleep sensor | 0 | 0 | 0 | 0 | 0 | 242 | 0 | 0 | 0 | 1 | 1 | 1 |
| 7- Netatmo weather station | 0 | 0 | 0 | 0 | 0 | 0 | 100 | 0 | 0 | 1 | 1 | 1 |
| 8- PIX-STAR photoframe | 0 | 0 | 7 | 0 | 0 | 0 | 0 | 626 | 0 | 1 | 0.989 | 0.994 |
| 9- Belkin Wemo motion sensor | 0 | 0 | 33 | 0 | 0 | 0 | 0 | 0 | 3434 | 1 | 0.99 | 0.995 |
| | | | | | | | | | **Weighted Avg** | **0.994** | **0.993** | **0.993** |

| Actual IoT device/classified as | 1 | 2 | 3 | 4 (U) | 5 | 6 | 7 | 8 | 9 | Precision | Recall | F1-Score |
|---|---|---|---|---|---|---|---|---|---|---|---|---|
| 1- Amazon Echo | 813 | 0 | 0 | 0 | 0 | 0 | 0 | 0 | 0 | 0.998 | 1 | 0.999 |
| 2- Samsung SmartCam | 0 | 323 | 0 | 0 | 0 | 0 | 0 | 0 | 0 | 0.883 | 1 | 0.938 |
| 3- Belkin Wemo switch | 0 | 0 | 365 | 0 | 0 | 0 | 0 | 0 | 0 | 1 | 1 | 1 |
| 4- Netatmo Welcome (unknown) | 2 | 43 | 0 | 259 | 0 | 1 | 0 | 0 | 2 | 0.814 | 0.844 | 0.829 |
| 5- Insteon camera | 0 | 0 | 0 | 0 | 210 | 0 | 0 | 0 | 0 | 1 | 1 | 1 |
| 6- Withings Aura smart sleep sensor | 0 | 0 | 0 | 0 | 0 | 242 | 0 | 0 | 0 | 0.996 | 1 | 0.998 |
| 7- Netatmo weather station | 0 | 0 | 0 | 0 | 0 | 0 | 100 | 0 | 0 | 1 | 1 | 1 |
| 8- PIX-STAR photoframe | 0 | 0 | 0 | 7 | 0 | 0 | 0 | 626 | 0 | 1 | 0.989 | 0.994 |
| 9- Belkin Wemo motion sensor | 0 | 0 | 0 | 52 | 0 | 0 | 0 | 0 | 3415 | 0.999 | 0.985 | 0.992 |
| | | | | | | | | | **Weighted Avg** | **0.985** | **0.983** | **0.984** |

| Actual IoT device/classified as | 1 | 2 | 3 | 4 | 5 (U) | 6 | 7 | 8 | 9 | Precision | Recall | F1-Score |
|---|---|---|---|---|---|---|---|---|---|---|---|---|
| 1- Amazon Echo | 812 | 0 | 0 | 0 | 1 | 0 | 0 | 0 | 0 | 0.985 | 0.999 | 0.992 |
| 2- Samsung SmartCam | 0 | 323 | 0 | 0 | 0 | 0 | 0 | 0 | 0 | 0.956 | 1 | 0.977 |
| 3- Belkin Wemo switch | 0 | 0 | 365 | 0 | 0 | 0 | 0 | 0 | 0 | 1 | 1 | 1 |
| 4- Netatmo Welcome | 0 | 0 | 0 | 306 | 1 | 0 | 0 | 0 | 0 | 1 | 0.997 | 0.998 |
| 5- Insteon camera (unknown) | 12 | 15 | 0 | 0 | 144 | 0 | 0 | 37 | 2 | 0.966 | 0.686 | 0.802 |
| 6- Withings Aura smart sleep sensor | 0 | 0 | 0 | 0 | 0 | 242 | 0 | 0 | 0 | 1 | 1 | 1 |
| 7- Netatmo weather station | 0 | 0 | 0 | 0 | 0 | 0 | 100 | 0 | 0 | 1 | 1 | 1 |
| 8- PIX-STAR photoframe | 0 | 0 | 0 | 0 | 1 | 0 | 0 | 626 | 6 | 0.944 | 0.989 | 0.966 |
| 9- Belkin Wemo motion sensor | 0 | 0 | 0 | 0 | 2 | 0 | 0 | 0 | 3465 | 0.998 | 0.999 | 0.999 |
| | | | | | | | | | **Weighted Avg** | **0.988** | **0.988** | **0.987** |

| Actual IoT device/classified as | 1 | 2 | 3 | 4 | 5 | 6 (U) | 7 | 8 | 9 | Precision | Recall | F1-Score |
|---|---|---|---|---|---|---|---|---|---|---|---|---|
| 1- Amazon Echo | 813 | 0 | 0 | 0 | 0 | 0 | 0 | 0 | 0 | 1 | 1 | 1 |
| 2- Samsung SmartCam | 0 | 323 | 0 | 0 | 0 | 0 | 0 | 0 | 0 | 1 | 1 | 1 |
| 3- Belkin Wemo switch | 0 | 0 | 365 | 0 | 0 | 0 | 0 | 0 | 0 | 1 | 1 | 1 |
| 4- Netatmo Welcome | 0 | 0 | 0 | 307 | 0 | 0 | 0 | 0 | 0 | 1 | 1 | 1 |
| 5- Insteon camera | 0 | 0 | 0 | 0 | 210 | 0 | 0 | 0 | 0 | 1 | 1 | 1 |
| 6- Withings Aura smart sleep sensor (unknown) | 0 | 0 | 0 | 0 | 0 | 239 | 0 | 0 | 3 | 1 | 0.988 | 0.994 |
| 7- Netatmo weather station | 0 | 0 | 0 | 0 | 0 | 0 | 100 | 0 | 0 | 1 | 1 | 1 |
| 8- PIX-STAR photoframe | 0 | 0 | 0 | 0 | 0 | 0 | 0 | 626 | 7 | 1 | 0.989 | 0.994 |
| 9- Belkin Wemo motion sensor | 0 | 0 | 0 | 0 | 0 | 0 | 0 | 0 | 3467 | 0.997 | 1 | 0.999 |
| | | | | | | | | | **Weighted Avg** | **0.998** | **0.998** | **0.998** |

| Actual IoT device/classified as | 1 | 2 | 3 | 4 | 5 | 6 | 7 (U) | 8 | 9 | Precision | Recall | F1-Score |
|---|---|---|---|---|---|---|---|---|---|---|---|---|
| 1- Amazon Echo | 813 | 0 | 0 | 0 | 0 | 0 | 0 | 0 | 0 | 1 | 1 | 1 |
| 2- Samsung SmartCam | 0 | 323 | 0 | 0 | 0 | 0 | 0 | 0 | 0 | 1 | 1 | 1 |
| 3- Belkin Wemo switch | 0 | 0 | 365 | 0 | 0 | 0 | 0 | 0 | 0 | 1 | 1 | 1 |
| 4- Netatmo Welcome | 0 | 0 | 0 | 306 | 0 | 0 | 1 | 0 | 0 | 1 | 0.997 | 0.998 |
| 5- Insteon camera | 0 | 0 | 0 | 0 | 210 | 0 | 0 | 0 | 0 | 1 | 1 | 1 |
| 6- Withings Aura smart sleep sensor | 0 | 0 | 0 | 0 | 0 | 242 | 0 | 0 | 0 | 1 | 1 | 1 |
| 7- Netatmo weather station (unknown) | 0 | 0 | 0 | 0 | 0 | 0 | 100 | 0 | 0 | 0.952 | 1 | 0.976 |
| 8- PIX-STAR photoframe | 0 | 0 | 0 | 0 | 0 | 0 | 1 | 626 | 6 | 1 | 0.989 | 0.994 |
| 9- Belkin Wemo motion sensor | 0 | 0 | 0 | 0 | 0 | 0 | 3 | 0 | 3464 | 0.998 | 0.999 | 0.999 |
| | | | | | | | | | **Weighted Avg** | **0.998** | **0.998** | **0.998** |

| Actual IoT device/classified as | 1 | 2 | 3 | 4 | 5 | 6 | 7 | 8 (U) | 9 | Precision | Recall | F1-Score |
|---|---|---|---|---|---|---|---|---|---|---|---|---|
| 1- Amazon Echo | 812 | 0 | 0 | 0 | 0 | 0 | 0 | 1 | 0 | 1 | 0.999 | 0.999 |
| 2- Samsung SmartCam | 0 | 323 | 0 | 0 | 0 | 0 | 0 | 0 | 0 | 1 | 1 | 1 |
| 3- Belkin Wemo switch | 0 | 0 | 364 | 0 | 0 | 0 | 0 | 1 | 0 | 1 | 0.997 | 0.999 |
| 4- Netatmo Welcome | 0 | 0 | 0 | 306 | 0 | 0 | 0 | 1 | 0 | 1 | 0.997 | 0.998 |
| 5- Insteon camera | 0 | 0 | 0 | 0 | 210 | 0 | 0 | 0 | 0 | 1 | 1 | 1 |
| 6- Withings Aura smart sleep sensor | 0 | 0 | 0 | 0 | 0 | 242 | 0 | 0 | 0 | 1 | 1 | 1 |
| 7- Netatmo weather station | 0 | 0 | 0 | 0 | 0 | 0 | 100 | 0 | 0 | 1 | 1 | 1 |
| 8- PIX-STAR photoframe (unknown) | 0 | 0 | 0 | 0 | 0 | 0 | 0 | 622 | 11 | 0.995 | 0.983 | 0.989 |
| 9- Belkin Wemo motion sensor | 0 | 0 | 0 | 0 | 0 | 0 | 0 | 0 | 3467 | 0.997 | 1 | 0.998 |
| | | | | | | | | | **Weighted Avg** | **0.998** | **0.998** | **0.998** |

| Actual IoT device/classified as | 1 | 2 | 3 | 4 | 5 | 6 | 7 | 8 | 9 (U) | Precision | Recall | F1-Score |
|---|---|---|---|---|---|---|---|---|---|---|---|---|
| 1- Amazon Echo | 810 | 0 | 0 | 0 | 0 | 0 | 0 | 0 | 3 | 1 | 0.996 | 0.998 |
| 2- Samsung SmartCam | 0 | 322 | 0 | 0 | 0 | 0 | 0 | 0 | 1 | 1 | 0.997 | 0.998 |
| 3- Belkin Wemo switch | 0 | 0 | 364 | 0 | 0 | 0 | 0 | 0 | 1 | 1 | 0.997 | 0.999 |
| 4- Netatmo Welcome | 0 | 0 | 0 | 306 | 0 | 0 | 0 | 0 | 1 | 1 | 0.997 | 0.998 |
| 5- Insteon camera | 0 | 0 | 0 | 0 | 209 | 0 | 0 | 0 | 1 | 1 | 0.995 | 0.998 |
| 6- Withings Aura smart sleep sensor | 0 | 0 | 0 | 0 | 0 | 242 | 0 | 0 | 0 | 1 | 1 | 1 |
| 7- Netatmo weather station | 0 | 0 | 0 | 0 | 0 | 0 | 100 | 0 | 0 | 1 | 1 | 1 |
| 8- PIX-STAR photoframe | 0 | 0 | 0 | 0 | 0 | 0 | 0 | 633 | 0 | 0.92 | 1 | 0.958 |
| 9- Belkin Wemo motion sensor (unknown) | 0 | 0 | 0 | 0 | 0 | 0 | 0 | 55 | 3412 | 0.998 | 0.984 | 0.991 |
| | | | | | | | | | **Weighted Avg** | **0.991** | **0.99** | **0.991** |